\documentstyle[prd,aps,preprint,epsfig]{revtex}
\tightenlines
\begin{document}
%%%%%%%%%%%%%%%%%%%%%%%%%%%%%%%%%%%%%%%%%%%%%%%%%%%%%%%%
\newcommand{\TeV}{\,{\rm TeV}}
\newcommand{\GeV}{\,{\rm GeV}}
\newcommand{\MeV}{\,{\rm MeV}}
\newcommand{\keV}{\,{\rm keV}}
\newcommand{\eV}{\,{\rm eV}}
\def\ap{\approx}
\def\bea{\begin{eqnarray}}
\def\eea{\end{eqnarray}}
\def\bec{\begin{center}}
\def\ec{\end{center}}
\def\pC{\tilde{\chi}^+}
\def\nC{\tilde{\chi}^-}
\def\pnC{\tilde{\chi}^{\pm}}
\def\Ne{\tilde{\chi}^0}
\def\snu{\tilde{\nu}}
\def\tN{\tilde N}
\def\ler{\lesssim}
\def\gtr{\gtrsim}
\def\beq{\begin{equation}}
\def\eeq{\end{equation}}
\def\haf{\frac{1}{2}}
\def\haf{\frac{1}{2}}
\def\plb#1#2#3#4{#1, Phys. Lett. {\bf #2B} (#4) #3}
\def\plbb#1#2#3#4{#1 Phys. Lett. {\bf #2B} (#4) #3}
\def\npb#1#2#3#4{#1, Nucl. Phys. {\bf B#2} (#4) #3}
\def\prd#1#2#3#4{#1, Phys. Rev. {\bf D#2} (#4) #3}
\def\prl#1#2#3#4{#1, Phys. Rev. Lett. {\bf #2} (#4) #3}
\def\mpl#1#2#3#4{#1, Mod. Phys. Lett. {\bf A#2} (#4) #3}
\def\rep#1#2#3#4{#1, Phys. Rep. {\bf #2} (#4) #3}
\def\lpp{\lambda''}
\def\ccg{\cal G}
\def\slash#1{#1\!\!\!\!\!/}
\def\rpv{\slash{R_p}}
\def\pslash{p\hspace{-2.0mm}/}
\def\qslash{q\hspace{-2.0mm}/}

%%%%% Song's Def %%%%%
\def\nub{\beta\beta_{0\nu}}
\def\sigu{\sigma^{\mu\nu}}
\def\sigd{\sigma_{\mu\nu}}
\def\bsigu{\bar{\sigma}^{\mu\nu}}
\def\bsigd{\bar{\sigma}_{\mu\nu}}
\newcommand{\imag}{\Im {\rm m}}
\newcommand{\real}{\Re {\rm e}}
\def\ph{arXiv:hep-ph/}
\def\th{arXiv:hep-th/}
%%%%%%%%%%%%%%%%%%%%%%

%%%%%%%%%%%%%%%%%%%%% Main Text %%%%%%%%%%%%%%%%
\setcounter{page}{1}
\draft
\preprint{KAIST-TH 02/17}
\title{Operator Analysis of Neutrinoless Double Beta Decay}
\author{Kiwoon Choi\footnote{kchoi@muon.kaist.ac.kr},
Kwang Sik Jeong\footnote{ksjeong@muon.kaist.ac.kr} 
and Wan Young Song\footnote{wysong@muon.kaist.ac.kr}}
\address{Department of Physics,
Korea Advanced Institute of Science and Technology\\
Daejeon 305-701, Korea}
\date{\today}
\maketitle
%
%%%%%%%%%%%%%%%%%%%%%%%%%%%%%%%%%%%%%%%%%%%%%%%%%%%%%%%%%%%%%
\begin{abstract}
%%%%%%%%%%%%%%%%%%%%%%%%%%%%%%%%%%%%%%%%%%%%%%%%%%%%%%%%%%%%%
We study the effective operators of the standard model fields
which would yield an observable rate of neutrinoless
double beta decay. We particularly focus on the possibility
that neutrinoless double beta decay is dominantly induced
by lepton-number-violating higher dimensional operators
other than the Majorana neutrino mass. 
Our analysis can be applied to models in which neutrinoless
double beta decay is induced either by a strong dynamics 
or by quantum gravity effects at a fundamental scale near the TeV scale
as well as the conventional models in which neutrinoless
double beta decay is induced by perturbative renormalizable interactions.
%%%%%%%%%%%%%%%%%%%%%%%%%%%%%%%%%%%%%%%%%%%%%%%%%%%%%%%%%%%%%
\end{abstract}
%%%%%%%%%%%%%%%%%%%%%%%%%%%%%%%%%%%%%%%%%%%%%%%%%%%%%%%%%%%%%
%
%
%%%%%%%%%%%%%%%%%%%%%%%%%%%
\section{Introduction}
\label{sec:intro}
%%%%%%%%%%%%%%%%%%%%%%%%%%%

The neutrinoless double beta decay ($\nub$) provides 
a very sensitive probe of lepton-number ($L$) violating interactions.
The most commonly quoted origin of $\nub$ is the
$ee$-component of Majorana neutrino mass matrix in the charged
lepton mass eigenbasis, 
which is given by
\bea
(m^\nu)_{ee}
= \sum_i U_{ei}^2 m_i
= c_{12}^2c_{13}^2e^{i\alpha_1}m_1+s_{12}^2
c_{13}^2e^{i\alpha_2}m_2+s_{13}^2e^{i\alpha_3}m_3\,,
\eea
where $m_i$ ($i=1,2,3$) are the neutrino mass
eigenvalues, $\theta_{ij}$ ($i\neq j$) and $\alpha_i$  
denote the mixing angles and CP phases in the
$3\times 3$ neutrino mixing matrix $U$, and
$c_{ij}=\cos\theta_{ij}$, $s_{ij}=\sin\theta_{ij}$.
Recently there has been a report to claim 
$\nub$ with half-life time 
$\tau_{1/2}\approx 10^{25}$ yrs \cite{exp}. 
Though the claim is still debatable~\cite{anti},  
some implications of this observation have been discussed already in
many papers \cite{after}.
If the origin of $\nub$  were due to $(m^\nu)_{ee}$, 
the data suggest
\bea
\label{numass_dbeta}
|(m^\nu)_{ee}|= 0.1 -0.6 \, ~{\rm eV}.
\eea
When combined with the information from atmospheric and solar
neutrino data, this value of $(m^\nu)_{ee}$ severely
constrains the possible form of neutrino mass matrix.
In particular, it does not allow the neutrino mass
eigenvalues in normal hierarchy.
As is well known, the atmospheric neutrino data imply~\cite{SK}
\bea
\Delta m^2_{\rm atm}
\equiv |m_3^2-m_2^2|\approx  3\times 10^{-3}\, {\rm eV}^2,\quad
\eea
As for the solar neutrino anomaly, the following four-solutions
are possible~\cite{BGP}:
\bea
&& \Delta m^2_{\rm sol}=|m_2^2-m_1^2| 
\quad\quad\quad \sin^2 2\theta_{12}
\nonumber \\
LMA: \quad\quad && \quad 3.2\times 10^{-5}\, {\rm eV}^2
\,\quad \quad\quad\quad \,\, 0.75\,
\nonumber \\
SMA: \quad\quad && \quad  5.0\times 10^{-6}\,
{\rm eV}^2\, \quad\quad\quad\quad 2.4\times 10^{-3}\,
\nonumber \\
LOW: \quad\quad && \quad 1.0\times 10^{-7} \, {\rm eV}^2\,
\quad\quad\quad\quad \,\,\,0.96 \,
\nonumber \\
VAC: \quad\quad && \quad 8.6\times 10^{-10} \, {\rm eV}^2\,
\quad\quad\quad\quad \,\,0.96
\eea
where LMA, SMA, LOW and VAC mean the large mixing angle MSW,
small mixing angle MSW, low mass, and vacuum oscillation solution,
respectively, and the numbers in each solution represent the best-fit values.
There exists also a constraint on $\theta_{13}$ from the
CHOOZ reactor experiment~\cite{CHOOZ}:
\beq
\sin\theta_{13} \,\lesssim\, 0.2\,.
\eeq
If the neutrino mass eigenvalues are in
normal hierarchy, i.e. $m_3\gg m_2 \gg m_1$,
which is one of  the plausible
scenarios, the above information on $\theta_{ij}$ and $m_i$ imply
\beq
\label{numass_osc}
|(m^\nu)_{ee}|\,=\,\left\{
\begin{array}{ll}
\,\, (2 \,-\, 5)\times 10^{-3} \, {\rm eV} & \quad\quad (LMA)
\\
\,\, 10^{-6} \, {\rm eV} \, -\, {\rm Max}(m_1, s_{13}^2 m_3)
& \quad\quad (SMA) \\
\,\, 10^{-4} \, {\rm eV} \, -\, s_{13}^2 m_3
& \quad\quad (LOW) \\
\,\, 10^{-5} \, {\rm eV}\, - \,  s_{13}^2 m_3
& \quad\quad (VAC)
\end{array} \right. 
\eeq
where 
\beq
s_{13}^2 m_3\,= \,2\times 10^{-3} \left(\frac{s_{13}^2}{4\times 10^{-2}}\right)
\, {\rm eV}\,\lesssim\, 2\times 10^{-3} \, {\rm eV}.
\eeq
Obviously, these values of $(m^\nu)_{ee}$ are too small to 
induce $\nub$ with $\tau_{1/2}\approx 10^{25}$ yrs.
So if  the claimed $\nub$ turns out to be correct,
we would have either (approximately) degenerate neutrino masses
or the observed $\nub$ is {\it not} due to 
$(m^\nu)_{ee}$, but due to some other $L$-violating 
interactions.
This would be true as long as 
$\tau_{1/2}\ll  10^{29}$ yrs which can be tested in future experiments
%CAMEO, CUORE, COBRA, ECHO, GENIUS, MAJORANA, MOON, XMASS etc~
\cite{exp2}.
%for instance in the MOON experiment which can probe
%$\nub$ up to $\tau_{1/2}\sim 10^{27}$ yrs or
%the GENIUS probing up to $\tau_{1/2}\sim 10^{29}-10^{30}$ yrs.

The possibility that $\nub$ is dominantly
induced by $L$-violating interactions {\it other than} 
$(m^\nu)_{ee}$  has been discussed before
in the context of specific models 
\cite{Rp,LR,tHiggs,vscal1,vscal2,diq1,lepto}, and also
a brief operator analysis of $\nub$  has been made in \cite{babu1}. 
In this paper, we wish to provide a more detailed operator
analysis of $\nub$ by studying
generic $L$-violating but baryon-number ($B$) conserving
higher-dimensional operators of the standard model (SM) fields
which may induce $\nub$.
The main focus will be given to the possibility
that $\tau_{1/2}(\nub)\approx 10^{25}$ yrs though
$(m^\nu)_{ee}$ is in the range of (\ref{numass_osc}).
Our result can be easily matched to the previous studies
on specific models in which $\nub$ is induced
by perturbative renormalizable interactions. 
It can be applied also to models
in which $\nub$ is induced  by either a strong dynamics
or quantum gravity effects at energy scales near the
TeV scale.

The organization of this paper is as follows.
In section II, we classify the $L$-violating operators
of the SM fields which can trigger $\nub$.
In section III, we tabulate the contraint on 
those $\Delta L=2$ operators 
from the condition $\tau_{1/2}(\nub)\gtrsim 10^{25}$ yrs,
and also estimate $(m^\nu)_{ee}$ which would be radiatively-induced 
by the operators triggering $\nub$.
In section IV,  we consider two specific 
models, i.e. a left-right symmetric model \cite{oriLR,vscal1}
and a model with scalar diquark and
dilepton \cite{diq2,diq1}, which can give  
$\tau_{1/2}(\nub)\approx 10^{25}$ yrs, while having $(m^\nu)_{ee}$
in the range of (\ref{numass_osc}).
We match our results to the previous analysis on these models. 
Section V is the conclusion.

%%%%%%%%%%%%%%%%%%%%%%%%%%%%%%%%%%%%%%%%%%%%%%%%%%%%%
\section{$L$-violating operators}
\label{sec:ops}
%%%%%%%%%%%%%%%%%%%%%%%%%%%%%%%%%%%%%%%%%%%%%%%%%%%%%

In this section, we classify the $L$-violating  but $B$-conserving
operators of the SM fields which would trigger $\nub$. 
A complete analysis of $\Delta L = 2$ operators which
would induce a  nonzero Majorana neutrino mass can be 
found in \cite{babu1}. The $\Delta L=2$  $\nub$ process may be induced 
by a double insertion of $\Delta L=1$ interactions or a single insertion
of $\Delta L=2$ interaction. However with the SM fields alone, there is 
no way to construct a $B$-conserving operator with $\Delta L=1$.
Since $\nub$ occurs at energy scales
far below the weak scale, the effects of quark-flavor-changing SM interactions 
on $\nub$ are suppressed by the small Fermi constant and also the small
quark-mixing angles. 
Also there is no renormalizable 
lepton-flavor-changing interaction in the SM. 
With these observations, we can ignore the effects of fermion-flavor-violation, so
limit the analysis to the $\Delta L=2$  operators
containing only the first generation of quarks and leptons.
We also limit our study only to the operators without spacetime
derivative or gauge field.

We use a notation in which all fermions are two-component Weyl 
spinors, i.e. $\psi$ is a left-handed spinor and $\bar{\psi}$ is its
right-handed hermitian conjugate. Generic fermion bilinear can be
a Lorentz scalar, vector, or tensor:
$$
(\psi\chi)_S=(\psi\chi),\quad
(\psi\bar{\chi})_V=(\psi\sigma^\mu\bar{\chi}),\quad
(\psi\chi)_T=(\psi\sigu\chi),
$$ 
where $\sigma_{\mu\nu} = (\sigma_{\mu}\bar{\sigma}_{\nu}
-\sigma_{\nu}\bar{\sigma}_{\mu})/4$.
Left-handed fermions relevant for $\nub$ are 
\bea
\ell =(1,2)_{-1/2}, \quad e^c=(1,1)_1, \quad
q =(3,2)_{1/6}, \quad
u^c =(\bar{3},1)_{-2/3}, \quad d^c=(\bar{3},1)_{1/3}\,,
\eea
where  $SU(3)_C \times SU(2)_L \times U(1)_Y$ quantum numbers are indicated
in parentheses.
The SM Higgs doublet is denoted by $H = (h^+, h^0)=(1,2)_{1/2}$ and 
$\bar{H}$ is its hermitian conjugate.
We then have the following $\Delta L=2$ lepton bilinears  
\bea
(\ell^i \ell^j)_S=(\ell^j\ell^i)_S,\quad
(\ell^i\ell^j)_T=-(\ell^j\ell^i)_T,\quad
(\ell^i\bar{e^c})_V,\quad
(\bar{e^c}\bar{e^c})_S,
\eea
where $i,j$ are $SU(2)$-doublet indices.
%In the following, we will keep  $SU(2)$-doublet indices 
%explicitly since it will be useful 
%for the dicussions of neutrino masses radiatively generated by
%$\Delta L=2$ operators.
% as will be given in Sec.\ref{sec:naive}.

There is a unique $\Delta L=2$, $\mbox{dimension} (d) = 5$ operator 
\bea
{\cal L}_{d=5} = 
  -\frac{\xi}{\Lambda}(\ell^i\ell^j)_S H^k H^l\epsilon_{ik}\epsilon_{jl} 
  + h.c. \,,
\label{eq:dim5}
\eea
where $\Lambda$ denotes the mass scale of
$L$-violating new physics which is assumed to exceed the weak scale, 
and the dimensionless $\xi$ represents the strength
of the couplings and/or the possible loop suppression factor
involved in the mechanism to generate the above $d=5$ operator.
After the electroweak symmetry breaking by $\langle H\rangle
=(0, v/\sqrt{2})$ with $v = 246$ GeV, it gives
the $ee$-component of the $3\times 3$ Majorana neutrino mass matrix:
\bea
(m^\nu)_{ee}=\frac{\xi v^2}{\Lambda} \,.
\eea
This neutrino mass is bounded to be less than 1 eV by $\nub$. 
Such a small neutrino mass can be a consequence
of very large value of  $\Lambda$ \cite{seesaw1,seesaw2}, 
e.g. $\xi\sim 1$ and $\Lambda\sim 10^{14}$ GeV.
Alternatively, $\Lambda$ can be of order TeV, but $m^\nu$ is small because
$\xi$ is small due to small couplings in the underlying 
dynamics which may be a consequence of some flavor symmetries \cite{choi}, 
e.g. $\xi\sim 10^{-11}$ and $\Lambda\sim 1$ TeV. 
At any rate, when combined with the double insertion of the
charged-current weak interaction, this Majorana neutrino mass leads to $\nub$
as in Fig. 1.a.

It is rather easy to see that there is no $B$-conserving
$d=6$, $\Delta L=2$ operator. 
As for $d=7$, $\Delta L=2$ operators which would trigger $\nub$, we have
\bea
{\cal L}_{d=7}=\frac{1}{\Lambda^3}&&\left[\,
\lambda_1^{S,T}\epsilon_{ik}\epsilon_{lj}
(\ell^i\ell^j)_{S,T}(q^k d^c)_{S,T} H^l 
+\lambda^T_2\epsilon_{ij}\epsilon_{kl}(\ell^i\ell^j)_T(q^kd^c)_TH^l
\right.\nonumber \\
&&\left.
+\lambda_3^{S,T}\epsilon_{jl}
(\ell^i\ell^j)_{S,T}(\bar{q_i}\bar{u^c})_{S,T} H^l 
+ \lambda^T_4\epsilon_{ij}(\ell^i\ell^j)_T(\bar{q}_k\bar{u^c})_TH^k 
\right.\nonumber \\
&&\left.
+\lambda_5 (\ell^i\bar{e^c})_V (d^c\bar{u^c})_V H^j \epsilon_{ij}
\,\right] 
+ h.c. \,,
\label{eq:dim7}
\eea
where again $\Lambda$ is the mass scale at which  the above operators are generated. 
After the electroweak symmetry breaking, these $d=7$
operators yield the following 4-fermion operators:
\bea
\frac{v}{\sqrt{2}\Lambda^3}&&\left[\,
  \lambda_1\,(\nu_e e)_S(u d^c)_S 
+ \lambda_2\,(\nu_ee)_T(ud^c)_T
+ \lambda_3\,(\nu_e e)_S(\bar{d}\bar{u^c})_S 
\right.\nonumber \\ 
&&\left.
+\lambda_4\,(\nu_ee)_T
(\bar{d}\bar{u^c})_T
+\lambda_5\,(\nu_e\bar{e^c})_V(d^c\bar{u^c})_V\,\right] 
+ h.c. \,,
\label{eq:dim7v}
\eea
where 
$$\lambda_1=\lambda^S_1,
\quad
\lambda_{2}=\lambda_2^T-\lambda_1^T,\quad
\lambda_3=\lambda^S_3,\quad
\lambda_4=\lambda_4^T-\lambda_3^T.
$$
When combined with a single insertion of the standard charged current
weak interaction,  
the above $\Delta L=2$ four-fermi interactions lead to $\nub$ as in Fig. 1.b.

As for the operators with $d\geq 8$, we are interested in the 
operators which can induce $\nub$ {\it without} involving the SM weak interaction.
Such operators should contain two electrons, so can be written as
\bea
{\cal L}_{d\geq 8} =
(\ell^i\ell^j)_{S}{\cal O}^{S}_{ij}
+(\bar{e^c}\ell^i)_V{\cal O}^V_i
+(\bar{e^c}\bar{e^c})_S{\cal O}^S\,
\eea
where ${\cal O}_I=\{{\cal O}^{S}_{ij}, {\cal O}^V_i,
{\cal O}^S\}$ are the operators made of the quarks and Higgs fields.
In order to be $\Delta Q_{\rm em}=2$, ${\cal O}_I$ must contain at least 4-quarks,
so we need $d\geq 9$.
Any $\Delta Q_{\rm em}=2$ four-quark operator can be written as a product
of two $\Delta Q_{\rm em}=1$ quark-antiquark bilinears:
${\cal O}_I=J_IJ_I^{\prime}$ where the quark-antiquark bilinears $J_I,
J_I^{\prime}$ can be either color-singlet or color-octet.
The hadronic matrix element for $\nub$ can be approximated as
$\langle Z+2|{\cal O}_I|Z\rangle\propto \langle p|J_I|n\rangle
\langle p|J^{\prime}_I|n\rangle$ for the neutron state $|n\rangle$ and
the proton state $|p\rangle$, and then $O_I$ with color-octet $J_I$
can be  ignored in the operator analysis of $\nub$.
The most general $d=9$ operators containing two color-singlet quark-antiquark bilinears 
together with two electrons
 are given by
\bea
{\cal L}_{d=9}=\frac{1}{\Lambda^5}&&\left[\,
\kappa_{1}^{S,T}\epsilon_{ik}\epsilon_{jl} 
(\ell^i\ell^j)_S (q^k d^c)_{S,T} (q^l d^c)_{S,T}
+\kappa_{2}^{S,T}
(\ell^i\ell^j)_S (\bar{q}_i\bar{u^c})_{S,T} (\bar{q}_j\bar{u^c})_{S,T}
\right.\nonumber \\
&&
 +\kappa_{3}^{S,T}\epsilon_{kj} 
(\ell^i\ell^j)_S (q^k d^c)_{S,T} (\bar{q}_i\bar{u^c})_{S,T}
+\kappa_{3}^{V}\epsilon_{ki}
(\ell^i\ell^j)_S (q^k\bar{q}_j)_V (d^c \bar{u}^c)_V
\nonumber \\
&&
+\kappa_4^{S,T}\epsilon_{ji} 
(\ell^i\bar{e^c})_V (d^c\bar{u^c})_V (q^j d^c)_{S,T}
+\kappa_5^{S,T}(\ell^i\bar{e^c})_V (d^c\bar{u^c})_V (\bar{q}_i \bar{u^c})_{S,T}
\nonumber \\
&&\left.
+\kappa_{6}(\bar{e^c}\bar{e^c})_S (d^c\bar{u^c})_V (d^c\bar{u^c})_V
\, \right] + h.c. \,.
\label{eq:dim9}
\eea
where all quark bilinears in the parentheses are color-
singlet.
%$(\ell^i\bar{e^c})_V (d^c\bar{u^c})_V (q^j d^c)_S =
%(\ell^i\sigma^{\mu}\bar{e^c}) (d^c\sigma_{\mu}\bar{u^c})(q^j d^c)$,
%$(\ell^i\bar{e^c})_V(d^c\bar{u^c})_V(q^jd^c)_T=
%(\ell^i\sigma^{\mu}\bar{e^c}) (d^c\sigma^{\nu}\bar{u^c})
%(q^j\sigd d^c)$, e.t.c.. 
These $d=9$ operators give the following 6-fermion operators
which would trigger $\nub$ as in Fig. 1.c:
\bea 
\frac{1}{\Lambda^5}&&\left[\,
 \kappa_{1}^{S,T}(ee)_S(u d^c)_{S,T}(u d^c)_{S,T}
+\kappa_{2}^{S,T}(ee)_S(\bar{d} \bar{u}^c)_{S,T}(\bar{d} \bar{u}^c)_{S,T}
\right. \nonumber   \\
&&\left.
+\kappa_{3}^{S,T}(ee)_S(u d^c)_{S,T}(\bar{d}\bar{u}^c)_{S,T}
+\kappa_{3}^V (ee)_S(u \bar{d})_V(\bar{u}^c d^c)_V
\right.\nonumber \\
&&\left.
+\kappa_4^{S,T}(e\bar{e}^c)_V(d^c\bar{u^c})_V(u d^c)_{S,T}
+\kappa_5^{S,T}(e\bar{e}^c)_V(d^c\bar{u^c})_V(\bar{d} \bar{u^c})_{S,T}
\right. \nonumber \\
&&\left.
+ \kappa_6(\bar{e^c}\bar{e^c})_S (d^c\bar{u^c})_V (d^c\bar{u^c})_V 
\, \right]+h.c.\,.
\label{eq:dim9b}
\eea

As for the next higher dimensional $d=11$ operators, 
we are interested in the operators which can {\it not}
be obtained by multiplying the gauge-invariant
Higgs bilnear $H^i\bar{H}_i$ to $d=9$ operators in (\ref{eq:dim9}).
Among such operators, the following ones are relevant for
$\nub$:
\bea
{\cal L}_{d=11}=\frac{1}{\Lambda^7}&&\left[\,
 \eta^{\prime}_{1}\epsilon_{km}\epsilon_{ln}
(\ell^i\ell^j)_S (q^k \bar{q}_i)_V (q^l \bar{q}_j)_V H^m H^n
+{\eta}^{\prime\prime}_1\epsilon_{kj}\epsilon_{ln}
(\ell^i\ell^j)_S(q^k\bar{q}_i)_V(q^l\bar{q}_m)_V H^mH^n
\right.\nonumber \\
&&
+\eta_1^{\prime\prime\prime}\epsilon_{ik}\epsilon_{jm}
(\ell^i\ell^j)_S(q^k\bar{q}_l)_V(q^m\bar{q}_n)_VH^lH^n
+\eta_{2} (\ell^i\ell^j)_S (d^c \bar{u^c})_V (d^c \bar{u^c})_V 
          \bar{H}_i \bar{H}_j
\nonumber \\
&&+\eta_{3}^{\prime S,T} \epsilon_{jl}\epsilon_{km}
(\ell^i\bar{e^c})_V (q^j \bar{q}_i)_V (q^k d^c)_{S,T} H^l H^m
+{\eta}^{\prime\prime,S,T}_3\epsilon_{ij}\epsilon_{ml}
(\ell^i\bar{e^c})_V(q^j\bar{q}_k)_V(q^l d^c)_{S,T}H^kH^m
\nonumber \\
&&+\eta^{\prime\prime\prime S,T}_3\epsilon_{il}\epsilon_{mj}
(\ell^i\bar{e}^c)_V(q^j\bar{q}_k)_V(q^l d^c)_{S,T}H^mH^k
+\eta^{\prime S,T}_{4}\epsilon_{jl} 
(\ell^i \bar{e^c})_V (q^j \bar{q}_i)_V (\bar{q}_k \bar{u^c})_{S,T} H^k H^l
\nonumber \\
&&+\eta^{\prime \prime S,T}_{4}\epsilon_{ji}(\ell^i\bar{e}^c)_V(q^j\bar{q}_k)_V(\bar{q}_l
\bar{u}^c)_{S,T}H^lH^k
+\eta^{\prime\prime\prime S,T}_4\epsilon_{jl}
(\ell^i\bar{e}^c)_V(q^j\bar{q}_k)_V(\bar{q}_i\bar{u}^c)_{S,T}H^lH^k
\nonumber \\
&&
+\eta_5^{S,T} \epsilon_{ik}\epsilon_{jl}
(\bar{e^c}\bar{e^c})_S (q^i d^c)_{S,T} (q^j d^c)_{S,T} H^k H^l 
+\eta_6^{S,T} (\bar{e^c}\bar{e^c})_S (\bar{q}_i \bar{u^c})_{S,T} 
(\bar{q}_j \bar{u^c})_{S,T} H^i H^j
\nonumber \\
&&\left.
+\eta_{7}^{S,T} \epsilon_{ik}
(\bar{e^c}\bar{e^c})_S (q^i d^c)_{S,T}(\bar{q}_j \bar{u^c})_{S,T} H^j H^k
+{\eta}_{7} \epsilon_{ik}
(\bar{e^c}\bar{e^c})_S (q^i \bar{q}_j)_V (d^c \bar{u^c})_V H^j H^k
\right],
\label{eq:dim11}
\eea
After the EWSB, these $d=11$ operators give the following
6-fermion operators leading to $\nub$ as in Fig. 1.c:
\bea
\frac{v^2}{2\Lambda^7}&&\left[\,
 \eta_1(ee)_S (u \bar{d})_V (u \bar{d})_V 
+\eta_2 (ee)_S (d^c \bar{u^c})_V (d^c \bar{u^c})_V
\right.\nonumber \\
&&\left.
+\eta_3^{S,T}(e\bar{e}^c)_V (u \bar{d})_V (u d^c)_{S,T}
+\eta_4^{S,T} (e\bar{e}^c)_V (u \bar{d})_V (\bar{d} \bar{u}^c)_{S,T}
\right. \nonumber \\
&&\left.
+\eta_5^{S,T} (\bar{e}^c \bar{e}^c)_S (u d^c)_{S,T} (u d^c)_{S,T}
+\eta_6^{S,T} (\bar{e}^c \bar{e}^c)_S 
              (\bar{d} \bar{u}^c)_{S,T} (\bar{d} \bar{u}^c)_{S,T}
\right. \nonumber \\
&&\left.
+\eta_{7}^{S,T} 
(\bar{e}^c \bar{e}^c)_S (u d^c)_{S,T} (\bar{d} \bar{u}^c)_{S,T} 
+\eta_{7} (\bar{e}^c \bar{e}^c)_S (u\bar{d})_V (d^c\bar{u}^c)_V
\, \right]+h.c\,, 
\label{eq:dim11v}
\eea
where
\bea
&& \eta_1=\eta^{\prime}_1+\eta^{\prime\prime}_1+\eta^{\prime\prime\prime}_1\,,
\nonumber \\
&&\eta_3^{S,T}=\eta^{\prime S,T}_3+{\eta}_3^{\prime\prime S,T}+
\eta^{\prime\prime\prime S,T}_3\,,
\nonumber \\
&& \eta^{S,T}_4=\eta^{\prime S,T}_4+\eta^{\prime\prime S,T}_4+
\eta^{\prime\prime\prime S,T}_4\,.
\eea

%
%%%%%%%%%%%%%%%%%%%%%%%%%%%%%%%%%%%%%%%%%%%%%%%%%%%%%%%%%%%%%%%%%%%%%%%%%%
\section{Constraints from $\nub$ }
\label{sec:naive}
%%%%%%%%%%%%%%%%%%%%%%%%%%%%%%%%%%%%%%%%%%%%%%%%%%%%%%%%%%%%%%%%%%%%%%%%%%

To determine the $\nub$ rate induced by the operators presented in section II, 
one needs to compute the nuclear matrix elements of the involved
multi-quark operators\footnote{In fact, since the $\Delta L=2$ operators are
assumed to be generated at scale $\Lambda$, one also needs to
compute the renormalization group evolution of the operators over the scales
from $\Lambda$ to $\Lambda_{QCD}\sim 1$ GeV.
Taking into account the effects of such renormalization group evolution
is beyond the scope of this paper, so will be ignored.}.
In this paper, we will use the results of \cite{super} for the necessary nuclear matrix
elements.
We will also assume that $\nub$ is dominated by one of the operators in
(\ref{eq:dim5}),(\ref{eq:dim7}), (\ref{eq:dim9})
and (\ref{eq:dim11}), so ignore possible interference
between the contributions from different operators.
The resulting $\tau_{1/2}$ have several sources of uncertainties, e.g. the RG evolution
effects, hadronic uncertainties in the nuclear matrix elements, and also
possible interference effects, however still it can be used to constrain 
$L$-violating interactions with a reasonable accuracy.

\begin{figure}
\begin{center}
\epsfig{file=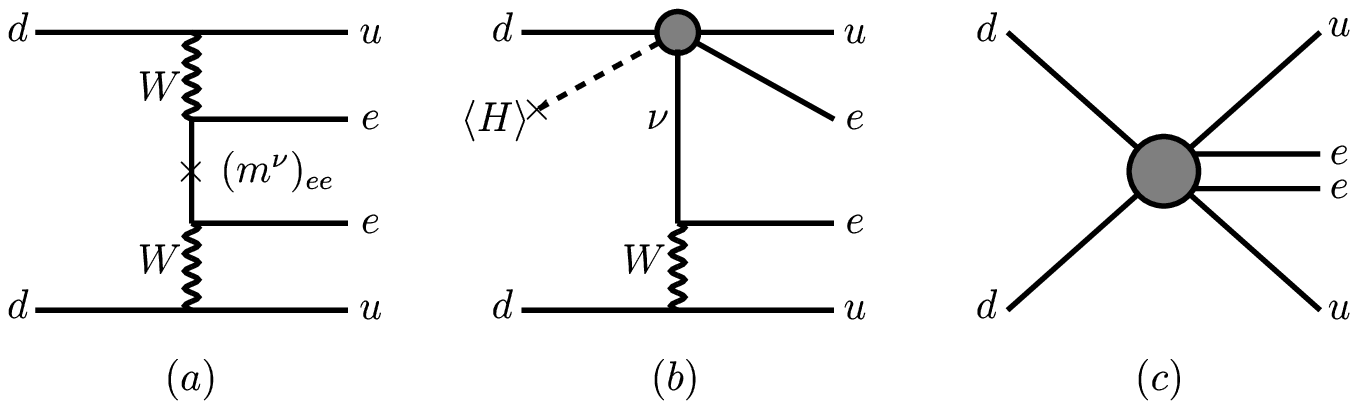,width=12cm,height=4cm}
\end{center}
\caption{
Feynman diagrams for $\nub$.
(a) corresponds to the conventional $\nub$ induced by $(m^\nu)_{ee}$.
(b) represents $\nub$ induced by the combined effects of
a $d=7$, $\Delta L=2$ operator (dark blob) and 
the SM charged current weak interaction, while (c)
represents $\nub$ induced by a $d=9$ or 11, $\Delta L=2$ operator.}
\label{fig1}
\end{figure}

If $\nub$ is induced dominantly by $(m^\nu)_{ee}$,
one finds \cite{exp}  
\bea
\tau_{1/2}^{-1} \,=
1.1\times 10^{-13}\left(\frac{v^2}{\Lambda m_e}\right)^2|\xi|^2 \,\,\mbox{yr}^{-1}
= \,7.4\times 10^{-30} \left(\frac{(m^{\nu})_{ee}}{4\times 10^{-3}\, {\rm eV}}
\right)^2 \,\, \mbox{yr}^{-1}
\label{eq:lifeMa}
\eea
In case that $\nub$ is dominated by one of $d\geq 7$ operators in section II,
the resulting $\tau^{-1}_{1/2}$ can be written
as $\tau_{1/2}^{-1} = |\epsilon|^2 \Phi_{\epsilon}$
where $\epsilon$ contains the operator coefficient, while
$\Phi_{\epsilon}$ contains the phase space factors and nuclear matrix elements
depending on the Lorentz structure of the corresponding operator.
Using the results of \cite{super},
the numerical values of $\Phi_{\epsilon}$ can be obtained as summarized in 
Tables \ref{tab:nuc7} and \ref{tab:nuc911}.
For $d=7$ operators of (\ref{eq:dim7}) giving the 4-fermion operators (\ref{eq:dim7v}), 
$\nub$ occurs as in Fig.\ref{fig1}(b).
We then find the corresponding half--life time
\bea
\tau_{1/2}^{-1} = \frac{1}{128}\left(\frac{v^3}{\Lambda^3}\right)^2
\left[~16 |\lambda_{1,3,5}|^2, |\lambda_{2,4}|^2\, \right]\Phi_{\lambda_i}\,\,\,\mbox{yr}^{-1},
\label{eq:longlife}
\eea
where the numerical values of $\Phi_{\lambda_i}$ are listed in Table \ref{tab:nuc7}.
The upperbound on $\lambda$'s resulting from
the condition $\tau_{1/2}\gtrsim 10^{25}$ yrs are summarized in 
Table~\ref{tab:7}.

It is also straightforward to compute
$\tau_{1/2}$ for $\nub$ induced by $d=9$ and $d=11$ operators
of (\ref{eq:dim9}) and (\ref{eq:dim11}).
For $d=9$ operators, we find
\bea
\tau_{1/2}^{-1} = \left(\frac{m_p v^4}{8 \Lambda^5}\right)^2
\left[~16 |\kappa^S_{1,2,3,4,5}|^2, 16 |\kappa^V_{3}|^2, 
   4 |\kappa^{T}_{4,5}|^2, |\kappa^{T}_{1,2,3}|^2, 16|\kappa_6|^2~\right]
~\Phi_{\kappa_i}\,\, \mbox{yr}^{-1},
\label{eq:short9}
\eea
and for $d=11$ operators, 
\bea
\tau_{1/2}^{-1} = \left(\frac{m_p v^6}{16 \Lambda^7}\right)^2
\left[~16 |\eta_{1,2,7}|^2, 16 |\eta^S_{3,4,5,6,7}|^2,
4 |\eta^T_{3,4}|^2, |\eta^T_{5,6,7}|^2~\right]
~\Phi_{\eta_i}\,\, \mbox{yr}^{-1},
\label{eq:short11} 
\eea
where $m_p$ is the proton mass and the numerical values of $\Phi_{\kappa_i}$
and $\Phi_{\eta_i}$ are listed in Table \ref{tab:nuc911}.
The resulting constraints on $\kappa$'s and $\eta$'s for
$\tau_{1/2}\gtrsim 10^{25}$ yrs are summarized in Table~\ref{tab:9} and
\ref{tab:11}.

The above equations (\ref{eq:longlife}),
(\ref{eq:short9}) and (\ref{eq:short11}) summarizing the $\nub$ rate
of $d=7,9,11$ operators and the resulting constraints on the operator
coefficients $\lambda$'s, $\kappa$'s and $\eta$'s listed
in Tables~\ref{tab:7},~\ref{tab:9},~\ref{tab:11}
are the main results of
this paper. 
Still one of our major concern is the possibility that 
$\nub$ is induced dominantly by one of $d\geq 7$ operators, not by
the $d=5$ operator for $(m^\nu)_{ee}$.
This would occur for instance if some of $\lambda$'s or $\kappa$'s saturate
their bounds from $\nub$, 
while $\xi$ is small enough to give $(m^\nu)_{ee}\ll 1$  eV.  
In fact, the condition $(m^\nu)_{ee}\ll 1$ eV constrains the coefficients of 
$d\geq 7$ operators also since those operators can generate 
$(m^\nu)_{ee}$ through loops.
For instance, the $d=7$ operator with coefficients $\lambda_{1,3}^S$ in (\ref{eq:dim7})
generate the $d=5$ operator for $(m^\nu)_{ee}$  through the one-loop diagram of   Fig. 2.a,
yielding
\bea
\delta_{\lambda}\xi \,\sim\, \frac{y_{d,u}}{16\pi^2} 
\lambda_{1,3}^S \,,
\label{eq:xil}
\eea
where we have taken the cutoff of the loop momenta
to be $\Lambda$ and $y_{d,u}$ is the down(up)--quark
Yukawa couplings. 
Other $d=7$ operators can generate $\xi$ also, however it involves 
more loops and/or more insertions of small Yukawa couplings.
For instance, the operator with coefficient $\lambda_5$ in (\ref{eq:dim7})
generates $\xi$ through the 2-loop diagram of Fig. 2.b, yielding
\bea
\delta_{\lambda}\xi\,\sim\, \frac{y_uy_dy_e}{(16\pi^2)^2}\lambda_5\,,
\eea
where $y_e$ is the electron Yukawa coupling.
As a result, $\delta_{\lambda} \xi$ from other $d=7$ operators are negligibly small
compared to $\delta_{\lambda}\xi$ from $\lambda^S_{1,3}$.

Similarly, the $d=9$ operators with coefficients 
$\kappa^{S}_{1,2,3}$ in (\ref{eq:dim9}) generate  the $d=5$ operator
for $(m^\nu)_{ee}$ at two-loop order (Fig. 2.c):
\bea
\delta_{\kappa}\xi \,\sim\, \frac{(y_d^2, y_u^2, y_d y_u)}{(16\pi^2)^2} 
\kappa^{S}_{1,2,3} \,,
\label{eq:xik}
\eea
where again the cutoff of the loop momenta is chosen to be $\Lambda$.
Other $d=9$ operators and also the $d=11$ operators
can generate $\xi$, however again the corresponding diagrams involve
more loops and/or more insertions of small Yukawa couplings.
For instance, the $d=9$ operator with coefficient $\kappa_6$ generates
$\xi$ through the 4-loop diagram of Fig. 2d, yielding
\bea
\delta_{\kappa}\xi \,\sim\, \frac{y_u^2y_d^2y_e^2}{(16\pi^2)^4}\kappa_6\,,
\eea
which is absolutely negligible even when $\kappa_6$ saturates the bound
from $\nub$.

Combining (\ref{eq:lifeMa}), (\ref{eq:longlife}) and (\ref{eq:short9}) with
(\ref{eq:xil}) and (\ref{eq:xik}), one easily finds
\bea
&&\frac{\tau^{-1}_{1/2}(\lambda^S_{1,3})}{\tau^{-1}_{1/2}(\delta_\lambda\xi)}\,\sim
\, 3\times 10^3 \left(\frac{\mbox{TeV}}{\Lambda}\right)^4\,,
\nonumber \\
&&\frac{\tau^{-1}_{1/2}(\kappa^S_{1,2,3})}{\tau^{-1}_{1/2}(\delta_\kappa\xi)}
\,\sim\, 3\times 10^7 \left(\frac{\mbox{TeV}}{\Lambda}\right)^8\,,
\eea
implying that if
the scale $\Lambda$ of $L$-violating interactions is about 1 TeV, 
it is quite possible that $\nub$ is dominantly induced by 
one of $\Delta L=2$, $d=7$ or $d=9$ operators.
In particular, one of $d=7$ or $d=9$ operators can give
$\tau_{1/2}(\nub)\approx 10^{25}$ yrs even when $(m^\nu)_{ee}$
is in the range of (\ref{numass_osc}) as
suggested by the atmospheric and solar neutrino
oscillation data in normal neutrino mass hierarchy.

%%%%%%%%%%%%%%%%%%%%%%%%%%%%%%%%%%%%%%%%%%%%%%%%%%%%%%%%%%%%%5
\section{Applications to some models}
\label{sec:app}
%%%%%%%%%%%%%%%%%%%%%%%%%%%%%%%%%%%%%%%%%%%%%%%%%%%%%%%%%%%%%%

Our results in the previous section can be applied to various kind of models 
providing $L$-violating interactions for $\nub$ and/or neutrino mass.
In this section, we consider some specific models of $\nub$ which
have been discussed in the literatures \cite{vscal1,diq1}
and use our results 
to rederive the constraints on $L$-violating couplings from the condition
$\tau_{1/2}(\nub)\gtr 10^{25}$ yrs.

Let us first consider a model in which  $d=7$ operators can be a 
dominant
source of $\nub$.
An example of such model is the left-right 
symmetric model~\cite{oriLR,vscal1} with gauge group 
$SU(3)_C \times SU(2)_L \times SU(2)_R \times U(1)_{B-L}$.
The Higgs sector of the model contains a bi-doublet $\phi$ and also
triplets ${\Delta}_L$ and ${\Delta}_R$ whose $SU(2)_L\times SU(2)_R\times U(1)_{B-L}$
quantum numbers are given by
\bea
&&\Delta_{L} = (3,1)_2=
\pmatrix{
\delta^+_L/\sqrt{2} & \delta^{++}_L \cr
\delta^0_L & -\delta^+_L/\sqrt{2} }\,,
\nonumber \\
&&\Delta_R=(1,3)_2=
\pmatrix{
\delta^+_R/\sqrt{2} & \delta^{++}_R\cr
\delta^0_R & -\delta^+_R/\sqrt{2}}\,,
\nonumber \\
&& \phi =(2,2)_0=\pmatrix{
\phi_1^0 & \phi_1^+ \cr \phi_2^- & \phi_2^0 } \,,
\nonumber
\eea
where the subscripts are the $U(1)_{B-L}$ charges.
The model contains also the left and right-handed lepton doublets:
\bea
\ell_L = \pmatrix{\nu \cr e} =(2,1)_{-1}, \quad
\bar{\ell}_R = \pmatrix{\bar{N}^c \cr \bar{e}^c}=(1,2)_{-1} \,,
\nonumber
\eea
as well as the quark doublets 
\bea
Q_L=\pmatrix{u\cr d}=(2,1)_{1/3},\quad
\bar{Q}_R=\pmatrix{\bar{u^c}\cr \bar{d^c}}=
(1,2)_{1/3}\,.
\nonumber
\eea
Yukawa interactions of the 1st generation  are given by
\bea
{\cal L}_Y &=& h\ell_{L} \phi \ell_{R} 
+ \tilde{h}\ell_{L}\sigma_2\phi^*\sigma_2\ell_{R}
+ h^{Q}Q_{L} \phi Q_{R}
+ \tilde{h}^{Q} Q_{L} \tilde{\phi} Q_{R}
\nonumber \\
&+& f \ell_{L} i\sigma_2\Delta_L  
\ell_{L} + f\ell_R i\sigma_2\Delta^{\dagger}_R \ell_R+h.c.  
\eea
Parameters in the Higgs potential can be chosen to yield
the Higgs vacuum expectation values
$\langle \delta^0_{L,R} \rangle =  v_{L,R}$ and 
$\langle \phi \rangle = \mbox{diag}(\kappa,~\kappa')$ with
the scale hierarchy
$v_R \gg \kappa\gg\kappa' \gg v_L$.
Then the fermion masses of the model are given by
$m_e\approx \tilde{h}\kappa$, $m^u\approx h^Q\kappa$,
$m^d\approx \tilde{h}^Q\kappa$ for $\kappa\sim 180$ GeV, and the 
neutrino mass
\bea
(m^\nu)_{ee}\approx \frac{(h\kappa)^2-4f^2v_Lv_R}{2fv_R}\,.
\eea
The model can generate also the $d=7$ operators of
(\ref{eq:dim7}) (with coefficients $\lambda^S_{1,3}$) through
the diagram of Fig.3.a, yielding
\bea
\frac{\lambda^S_1}{\Lambda^3}&\approx& \frac{\gamma v_R f}
{m_{\Delta_L}^2}\left(
\frac{h^Q}{m_{\phi_1}^2}-\frac{\tilde{h}^Q}{m_{\phi_2}^2}\right)\,,
\nonumber \\
\frac{\lambda^S_3}{\Lambda^3}&\approx&\frac{\gamma v_R f}{m_{\Delta_L}^2}
\left(\frac{\tilde{h}^Q}{m_{\phi_1}^2}-\frac{h^Q}{m_{\phi_2}^2}\right)\,,
\label{eq:matching}
\eea
where $\gamma$ is the coefficient of the term 
$\mbox{tr}(\Delta^{\dagger}_L\phi\Delta_R\phi^{\dagger})$ 
in the Higgs potential.
Then there exists a parameter range of the model in which
$\nub$ is dominated by these $d=7$ operators.
For instance, if $fv_R\sim m_{\Delta_L}\sim 10^5$ GeV, 
$m_{\phi}\sim 2\times 10^2$ GeV, and $\gamma\sim 10^{-1}$, the resulting
$\lambda^S_{1,3}/\Lambda^3\sim
10^{-6}/(\mbox{TeV})^3$ saturates the bound listed in Table~\ref{tab:7},
so lead to $\nub$ with
$\tau_{1/2}\sim 10^{25}$ yrs.
Though not very natural, still
the parameters of the model can be tuned to yield
$(m^\nu)_{ee}={\cal O}(10^{-3})$ eV, while keeping
$\lambda^S_{1,3}/\Lambda^3\sim 10^{-6}/(\mbox{TeV})^3$.
So the model can accomodate
$\tau_{1/2}(\nub)\sim 10^{25}$ yrs
together with the atmospheric and solar neutrino
oscillation data in hierarchical neutrino mass scenario.

As an example of model in which $d=9$ operators of
(\ref{eq:dim9}) can be a dominant source of $\nub$,
let us consider a model containing scalar-diquarks and 
scalar-dilepton \cite{diq2,diq1}
with the following $SU(3)_c\times SU(2)_L\times U(1)_Y$
quantum numbers: 
\bea
\Delta_u =(6,1)_{4/3}, \quad \Delta_d =(6,1)_{-2/3}, 
\quad \Delta_e =(1,1)_{-2}\,.
\eea
The Yukawa couplings of the model are assumed to include
\bea 
  h_u \Delta_u u^c u^c
+ h_d \Delta_d d^cd^c
+ h_e \Delta_e e^c e^c 
\label{eq:dil}
\eea
and also the Higgs potential contains
\bea
\mu \Delta_u\Delta_d^{\dagger}\Delta_e
\label{eq:diq}
\eea
which breaks $L$-conservation.

With the above interactions,
a $d=9$ operator for $\nub$ is generated at tree level as depicted in
Fig. 3b. The resulting operator corresponds to the $\kappa_6$ term of (\ref{eq:dim9}):
\bea
\frac{\kappa_6}{\Lambda^5}(\bar{e^c}\bar{e^c})_S (d^c\bar{u^c})_V (d^c\bar{u^c})_V\,,
\label{eq:diqe}
\eea
where
\bea
\frac{\kappa_6}{\Lambda^5}=\frac{\mu h_u h_d h_e}{8 m^2_{\Delta_u} m^2_{\Delta_d} m^2_{\Delta_e}}.
\eea
If $h_{u,d,e}\sim 1$, 
$m_{\Delta_{u,d,e}} \sim 1$ TeV
and $\mu \sim 250$ GeV, the resulting 
$\kappa_6/\Lambda^5\sim 3\times 10^{-2}/(\mbox{TeV})^5$
saturates the bound in Table IV, so the model leads to $\nub$ 
with $\tau_{1/2} \sim 10^{25}$ yr. 
On the other hand, $(m^\nu)_{ee}$ induced by the $L$-violating
interaction (\ref{eq:diq}) is 4-loop suppressed and involves
6-powers of small Yukawa couplings as in Fig. 2.d:
\bea
(m^\nu)_{ee}\,\sim\,
\frac{y_u^2y_d^2y_e^2}{(16\pi^2)^4}\frac{h_uh_dh_e \mu
\langle H\rangle^2}{m^2_{\Delta}}.
\eea
This $(m^\nu)_{ee}$ is absolutely negligible, $(m^\nu)_{ee}\sim 10^{-30}$ eV,
even when $\kappa_6/\Lambda^5$ saturates its bound.
So one needs additional $L$-violating interactions to
generate neutrino masses which would explain the atmospheric and 
solar neutrino oscillation data in hierarchical neutrino mass scenario.

%
%%%%%%%%%%%%%%%%%%%%%%%%%%%%%%%%%%%%%%%%%%%%%%%%%%%%%%%%%%%%%5
\section{Conclusion}
%%%%%%%%%%%%%%%%%%%%%%%%%%%%%%%%%%%%%%%%%%%%%%%%%%%%%%%%%%%%%%

Motivated by the recent claim of observing $\nub$ with $\tau_{1/2}
\approx 10^{25}$ yrs, we studied the effective $\Delta L=2$ operators of the SM fields
which would generate $\nub$.
We classified such operators up to mass dimension $d=11$, and
find the upper bound on each operator coefficient resulting from
the condition $\tau_{1/2} \gtrsim 10^{25}$ yrs. Our results
are summarized in Tables III, IV, V.
We also examined the possibility that $d=7$ or $9$ operators
are a dominant source of $\nub$  in the
context of generic operator analysis, particularly the possibility 
that $\tau_{1/2}(\nub)\sim 10^{25}$ yrs, while $(m^\nu)_{ee}={\cal O}(10^{-3})$ eV as suggested by
the atmospheric and solar neutrino oscillation data in
hierarchical neutrino mass scenario.
As we have demonstrated in section IV, our results can be easily matched 
to the previous analysis on
specific models in which $\nub$ is induced by perturbative renormalizable
interactions. They can be also applied to models
in which $\nub$ is induced by either a strong dynamics or quantum gravity
effects at scales near the TeV scale.

\bigskip
%%%%%%%%%%%%%%%%%%%%%%%%%%%%%%%%%%%%%%%%%%%%%%%%%%%%%%%%%
\acknowledgements
%%%%%%%%%%%%%%%%%%%%%%%%%%%%%%%%%%%%%%%%%%%%%%%%%%%%%%%%%

This work is supported in part by BK21 Core program of MOE, 
KRF Grant No. 2000-015-DP0080,
KOSEF Sundo-Grant, and KOSEF
through CHEP at Kyungpook National University.
%
%%%%%%%%%%%%%%%%%%%%%%%%%%%%%%%%%%%%%%%%%%%%%%%%%%%%%%%%%%%

%%%%%%%%%%%%%%%%%%%%%%%%%%%%%%%%%%%%%%%%%%%%%%%%%%%%%%%%%%%
%
%%%%%%%%%%%%%%%%%%%%%%%
% Figure
%%%%%%%%%%%%%%%%%%%%%%%
%
%
%\bigskip
%
\begin{figure}
\begin{center}
\epsfig{file=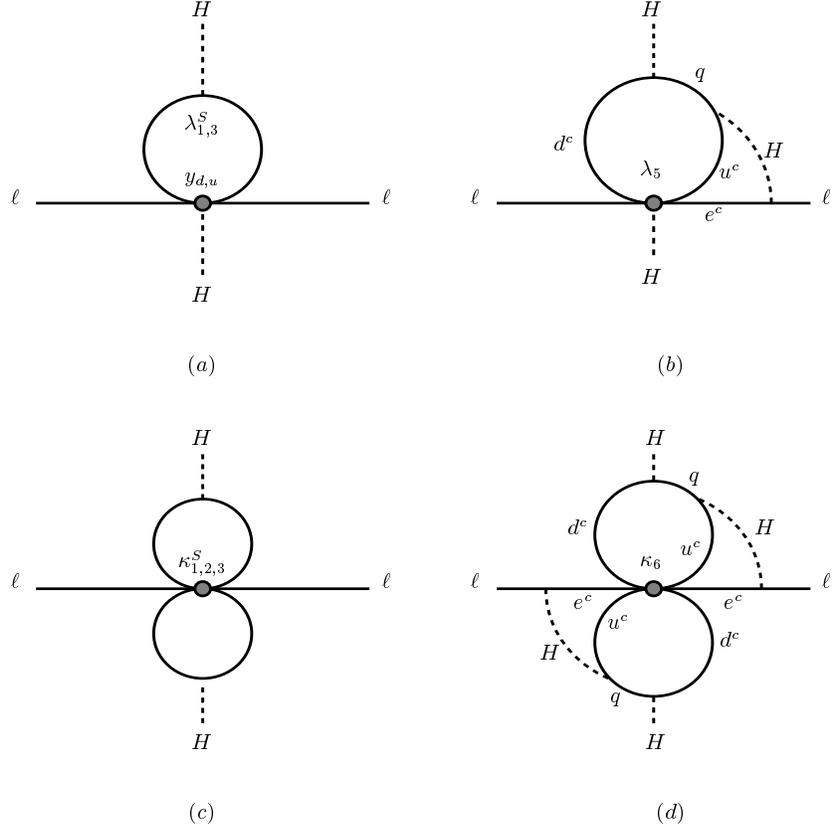,width=11cm,height=11cm}
\end{center}
\caption{
Feynman diagrams for the $d=5$ operator for $(m^\nu)_{ee}$ 
radiatively induced by $d=7$ or $d=9$ operators
}
\label{fig2}
\end{figure}
%
%\bigskip
%
\begin{figure}
\begin{center}
\epsfig{file=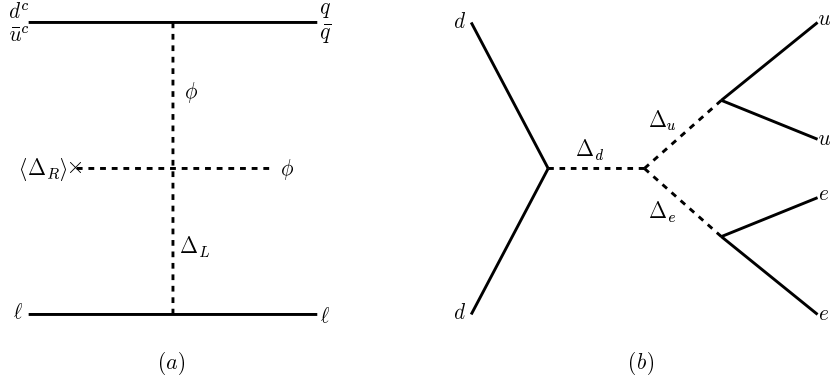,width=11cm,height=5cm}
\end{center}
\caption{
(a) Feynman diagram for the $d=7$ operator with coefficients
$\lambda^S_{1,3}$ in the left--right
symmetric model. 
(b) Feynman diagram for the $d=9$ operator with coefficient
$\kappa_6$ induced by the exchange of
scalar diquarks and dilepton.
}
\label{fig3}
\end{figure}
%
%%%%%%%%%%%%%%%%%%%%%%%%%%%%%%%%%
% TABLES
%%%%%%%%%%%%%%%%%%%%%%%%%%%%%%%%%
%
\def\arraystretch{1.5}
\begin{table}
\caption{Numerical values of $\Phi_{\lambda_i}$ obtained using 
the results of Ref.\protect\cite{super}.}
\label{tab:nuc7}
\vspace{0.2cm}
\begin{center}
\begin{tabular}{|c|c|c|c|c|}
$\lambda_1$ & $\lambda_2$ & $\lambda_3$ & $\lambda_4$
            & $\lambda_5$ \\
\hline
$6.9\times 10^{-10}$ & $2.9\times 10^{-8}$ & $6.9\times 10^{-10}$ &
$2.0\times 10^{-7}$ & $1.7\times 10^{-13}$  \\
\end{tabular}
\end{center}
\end{table}
\vspace{-1cm}
\begin{table}
\caption{Numerical values of $\Phi_{\kappa_i}$ and $\Phi_{\eta_i}$ 
obtained using the results of Ref. \protect\cite{super}.}
\label{tab:nuc911}
\vspace{0.2cm}
\begin{center}
\begin{tabular}{|c|c|c|c|c|c|}
$\kappa^S_{1,2,3}, \eta^S_{5,6,7}$ & 
$\kappa^T_{1,2,3}, \eta^T_{5,6,7}$ & 
$\kappa_6, \eta_{1,2}$ & $\kappa_{3}, \eta_{7}$
            & $\kappa^S_{4,5}, \eta^S_{3,4}$ 
            & $\kappa^T_{4,5}, \eta^T_{3,4}$ \\
\hline
$6.2\times 10^{-13}$ & $1.4\times 10^{-8}$ & $3.5\times 10^{-11}$ &
$5.6\times 10^{-10}$ & $1.4\times 10^{-12}$ & $1.4\times 10^{-10}$ \\
\end{tabular}
\end{center}
\end{table}
\vspace{-1cm}
\begin{table}
\caption{Upperbounds on the coefficient of $d=7$ operators from 
$\tau_{1/2}\geq 10^{25}$ yrs. Here $\Lambda_{TeV}=\Lambda/\mbox{TeV}$} 
\label{tab:7} 
\vspace{0.2cm}
\begin{center}
\begin{tabular}{|c|c|c|c|c|}
$\lambda_1/\Lambda_{\rm TeV}^3$ & $\lambda_2/\Lambda_{\rm TeV}^3$ & 
$\lambda_3/\Lambda_{\rm TeV}^3$ & $\lambda_4/\Lambda_{\rm TeV}^3$ & 
$\lambda_5/\Lambda_{\rm TeV}^3$ \\
\hline
  $2.3\times 10^{-6}$ & $1.4\times 10^{-6}$ & $2.3\times 10^{-6}$
& $5.3\times 10^{-7}$ & $1.4\times 10^{-4}$ \\
\end{tabular}
\end{center}
\end{table}
\vspace{-1cm}
\begin{table}
\caption{Upperbounds on the coefficients of $d=9$ operators}
\label{tab:9}
\vspace{0.2cm}
\begin{center}
\begin{tabular}{|c|c|c|c|c|c|}
  $\kappa^S_{1,2,3}/\Lambda_{\rm TeV}^5$ & 
  $\kappa^T_{1,2,3}/\Lambda_{\rm TeV}^5$ & 
  $\kappa_{3}^V/\Lambda_{\rm TeV}^5$ & 
  $\kappa^S_{4,5}/\Lambda_{\rm TeV}^5$ & 
  $\kappa^T_{4,5}/\Lambda_{\rm TeV}^5$ & 
  $\kappa_6/\Lambda_{\rm TeV}^5$\\
\hline
  $2.3\times 10^{-1}$
& $6.2\times 10^{-3}$ & $7.7\times 10^{-3}$
& $1.5\times 10^{-1}$ & $3.1\times 10^{-2}$ 
& $3.1\times 10^{-2}$\\
\end{tabular}
\end{center}
\end{table}
\vspace{-1cm}
\begin{table}
\caption{Upper bounds on the coefficients of $d=11$ operators}
\label{tab:11}
\vspace{0.2cm}
\begin{center}
\begin{tabular}{|c|c|c|c|c|c|}
  $\eta_{1,2}/\Lambda_{\rm TeV}^7$ & 
  $\eta^S_{3,4}/\Lambda_{\rm TeV}^7$ & 
  $\eta^T_{3,4}/\Lambda_{\rm TeV}^7$ & 
  $\eta^S_{5,6,7}/\Lambda_{\rm TeV}^7$ & 
  $\eta^T_{5,6,7}/\Lambda_{\rm TeV}^7$ & 
  $\eta_{7}/\Lambda_{\rm TeV}^7$ \\
\hline
  $1.0$
& $5.1$ & $1.0$ & $7.6$
& $0.2$ & $0.3$ \\
\end{tabular}
\end{center}
\end{table}
%
%%%%%%%%%%%%%%%%%%%%%%%%%%%%%%%%%
\end{document}